\documentclass{ws-procs9x6}

\newcommand{\CK}{\v Cerenkov}

\begin{document}

\title{The RICH detector of the AMS-02 experiment: status and physics prospects}

\author{\underline{RUI PEREIRA}, on behalf of the AMS RICH collaboration}
\address{LIP/IST \\
         Av. Elias Garcia, 14, 1$^o$ andar, 1000-149 Lisboa, Portugal \\
         e-mail: pereira@lip.pt}

\vspace{-0.3cm}

\begin{abstract}
The Alpha Magnetic Spectrometer (AMS), whose final version AMS-02 is to be
installed on the International Space Station (ISS) for at least 3 years,
is a detector designed to measure charged cosmic ray spectra with energies
up to the TeV region and with high energy photon detection capability up
to a few hundred GeV. It is equipped with several subsystems, one of which
is a proximity focusing RICH detector with a dual radiator (aerogel+NaF)
that provides reliable measurements for particle velocity and charge. The
assembly and testing of the AMS RICH 
is currently being finished 
and the full AMS detector is expected to be ready by the end of 2008. The
RICH detector of AMS-02 is presented. Physics prospects are briefly
discussed.
\end{abstract}

\bodymatter

\vspace{-0.3cm}

\section{The AMS-02 experiment}

The Alpha Magnetic Spectrometer (AMS)\cite{bib:ams}, whose final version
AMS-02 is to be installed on the International Space Station (ISS) for at
least 3 years, is a detector designed to study the cosmic ray flux by
direct detection of particles above the Earth's atmosphere, at an
altitude of \mbox{$\sim$ 400 km}, using
state-of-the-art particle identification techniques. AMS-02 is equipped
with a superconducting magnet cooled by superfluid helium. The
spectrometer is composed of several subdetectors: a Transition Radiation
Detector (TRD), a Time-of-Flight (ToF) detector, a Silicon Tracker,
Anticoincidence Counters (ACC), a Ring Imaging \CK\ (RICH) detector and an
Electromagnetic Calorimeter (ECAL). A preliminary version of the
detector, AMS-01, was successfully flown aboard the US space shuttle
Discovery in June 1998.

\begin{figure}[htb]

\center

\vspace{-0.5cm}

\begin{tabular}{cc}  

\mbox{\epsfig{file=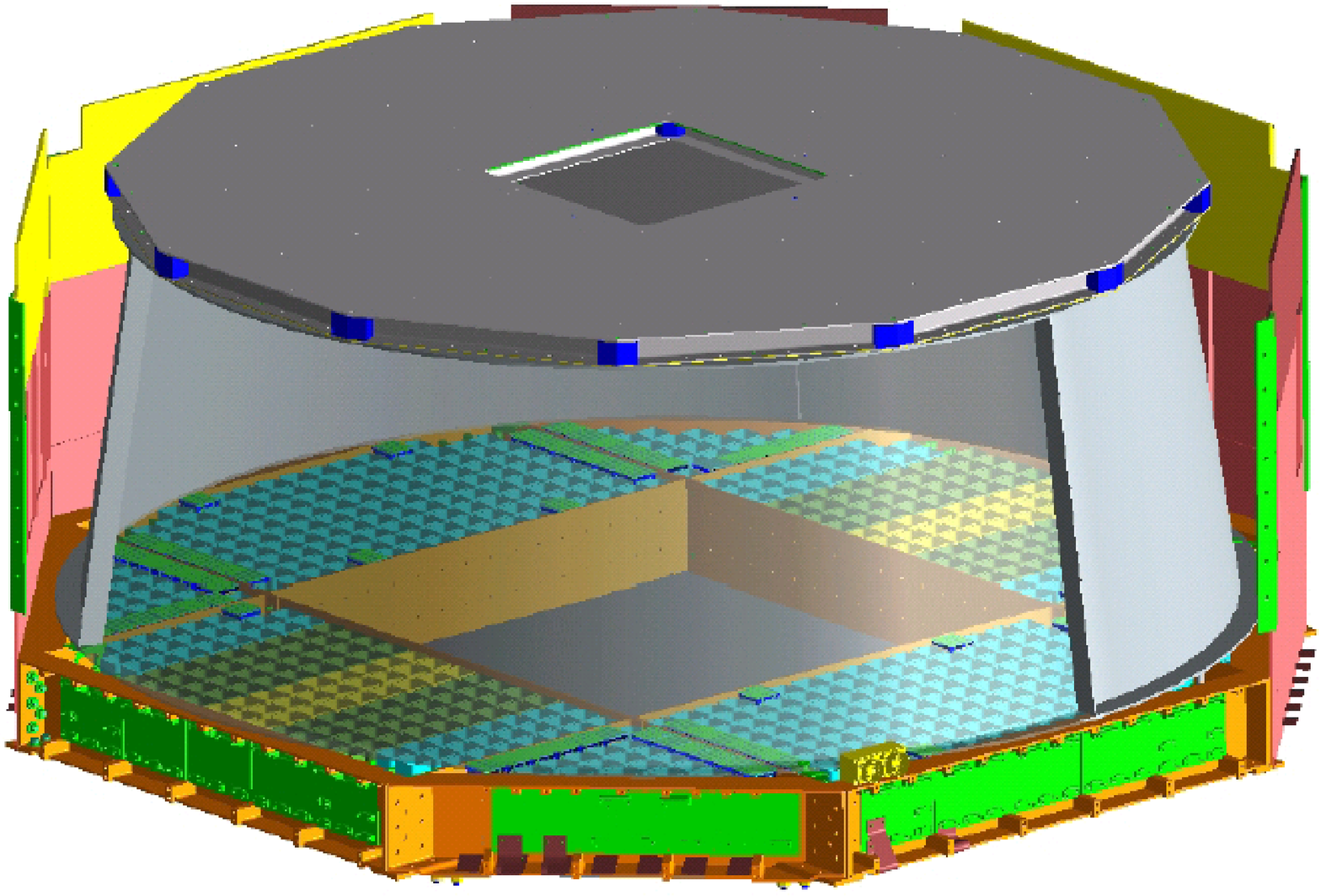,width=0.47\textwidth,clip=}}
&
\mbox{\epsfig{file=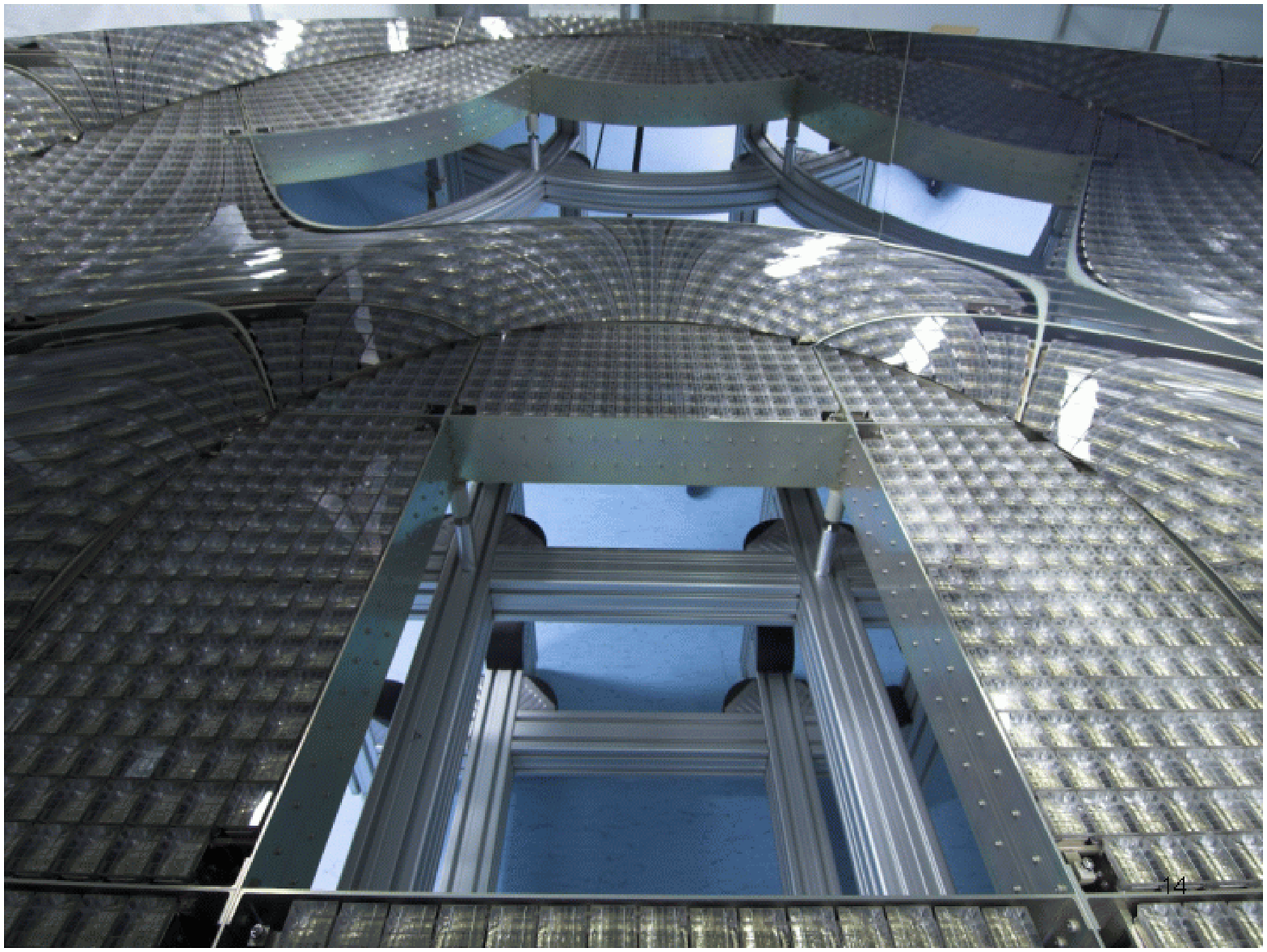,width=0.43\textwidth,clip=,
bbllx=-53,bblly=100,bburx=667,bbury=666}}

\end{tabular}

\vspace{-0.2cm}

\caption{The RICH detector of AMS-02 \emph{(left)}. View of
the assembled RICH detector at CIEMAT \emph{(right)}.\label{richdet}}

\vspace{-0.2cm}

\end{figure}

The main goals of the AMS-02 experiment are: (i) a precise measurement of
charged cosmic-ray spectra in the rigidity region between
\mbox{$\sim$ 0.5 GV} and \mbox{$\sim$ 2 TV} and the detection of photons
with energies up to a few hundred GeV; (ii) a search for heavy antinuclei
\mbox{($Z \ge$ 2)}, which if discovered would signal the existence of
cosmological antimatter; (iii) a search for dark matter constituents by
examining possible signatures of their presence in cosmic ray spectra.
The long exposure time and large acceptance \mbox{(0.5 m${}^2\cdot$sr)} of
AMS-02 will enable it to collect an unprecedented statistics of more than
$10^{10}$ nuclei.

\vspace{-0.3cm}

\section{The AMS RICH detector}

AMS-02 includes a proximity focusing Ring Imaging
\CK\ (RICH) detector, shown in Fig.~\ref{richdet}, placed in
the lower part of the spectrometer between the lower ToF counters and the
ECAL. A dual radiator configuration with silica aerogel ($n=$~1.050) and
sodium fluoride (NaF, $n=$~1.334) has been chosen for the RICH. The
geometrical acceptance of the central NaF square corresponds to
\mbox{$\sim$ 10\%} of the total RICH acceptance. 

A high reflectivity ($>85\%$ at \mbox{$\lambda = 420$ nm}) lateral conical
mirror made of aluminium-nickel-coated
silica has been included to increase photon collection. The detection
matrix at the bottom of the detector has 680 multianode photomultiplier
tubes (PMTs)
\mbox{(4 $\times$ 4)} coupled to light guides with a pixel size of 8.5 mm.

Charged particles crossing a radiator of refractive index $n$ with a
velocity $v>c/n$ emit \CK\ radiation in a cone with an aperture
$\theta_c = \arccos (\frac{1}{\beta n})$. The intensity of radiation is
proportional to the square of particle charge and also increases with velocity.
The impact of photons in the PMT matrix produces a ring which, combined with
data on particle tracks obtained from the Silicon Tracker, allows to
determine the \CK\ angle $\theta_c$ and therefore the particle's velocity.
Particle charge may be determined from the total signal collected by the
PMTs taking ring acceptance into account.

The analysis of RICH data involves the identification of the \CK\ ring in
a hit pattern which usually includes several scattered noise hits and an
eventual strong spot in the region where the charged particle crosses the
detection plane.

The RICH detector will provide a very accurate velocity measurement (in
aerogel, $\Delta \beta / \beta \sim$ 10${}^{-3}$ and 10${}^{-4}$ for
\mbox{$Z=$~1} and \mbox{$Z=$~10~$-$~20}, respectively) and charge
identification of nuclei up to iron \mbox{($Z=$~26)}.
RICH data, combined with information on particle rigidity from the AMS
Silicon Tracker, enable the reconstruction of particle mass. The accuracy
of the RICH velocity measurement is essential for mass reconstruction due
to the growth of relative errors when $\beta \sim 1$
$( \frac{\Delta m}{m} = \frac{\Delta p}{p} \oplus \gamma^2 \frac{\Delta
\beta}{\beta} )$.
The RICH detector also plays a major role in the exclusion of upward going
particles (albedo). Additional information on the RICH detector
may be found in Ref. \refcite{bib:icrc2005}.
The assembly of the AMS RICH detector is coming to an end at CIEMAT
in Madrid. The integration of the RICH and the other subdetectors of AMS-02
will take place at CERN and should be finished by the end of 2008.

\begin{figure}[htb]

\center

\vspace{-0.9cm}

\begin{tabular}{cc}  

\mbox{\epsfig{file=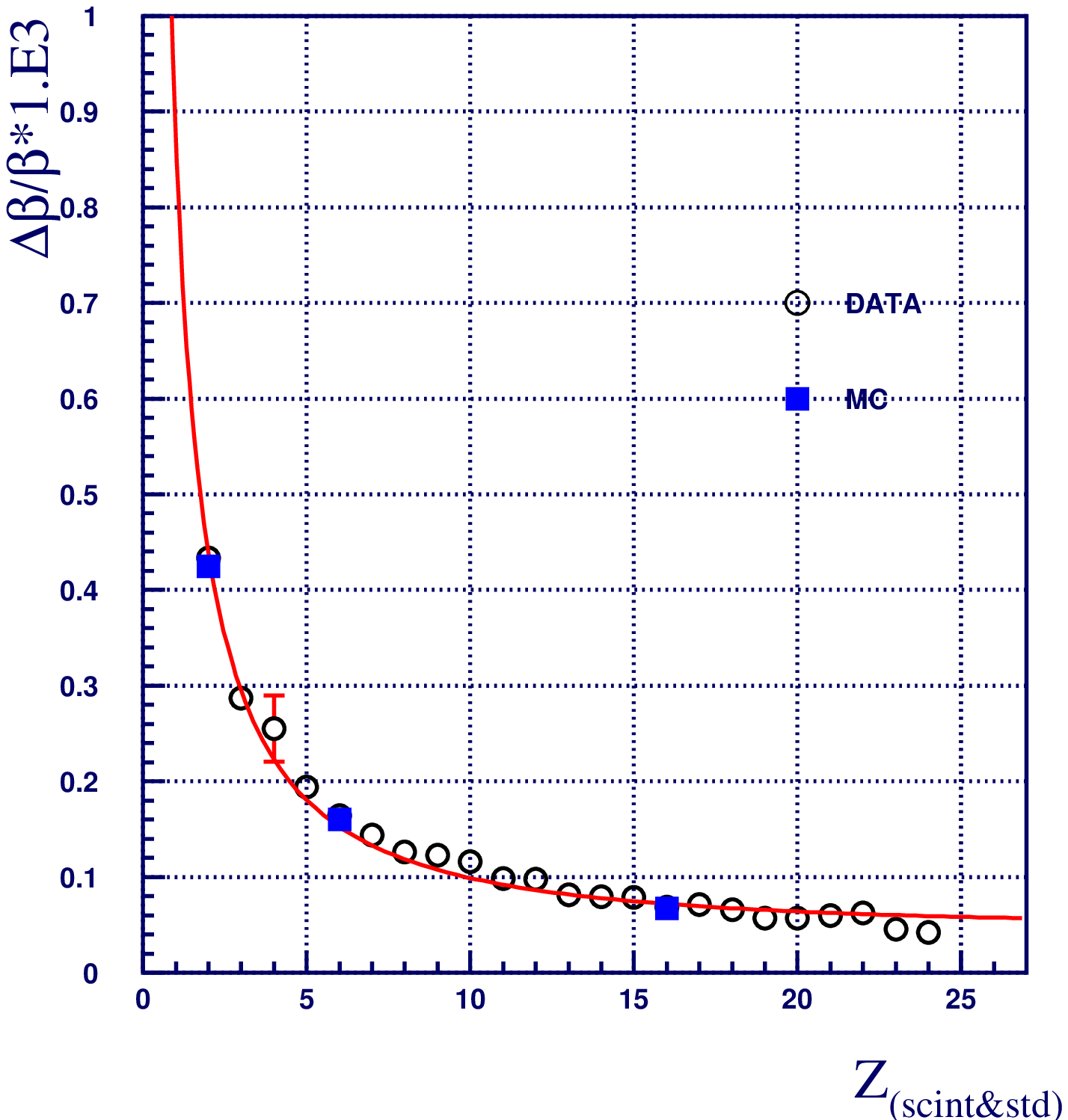,width=0.5\textwidth,clip=}}
&
\mbox{\epsfig{file=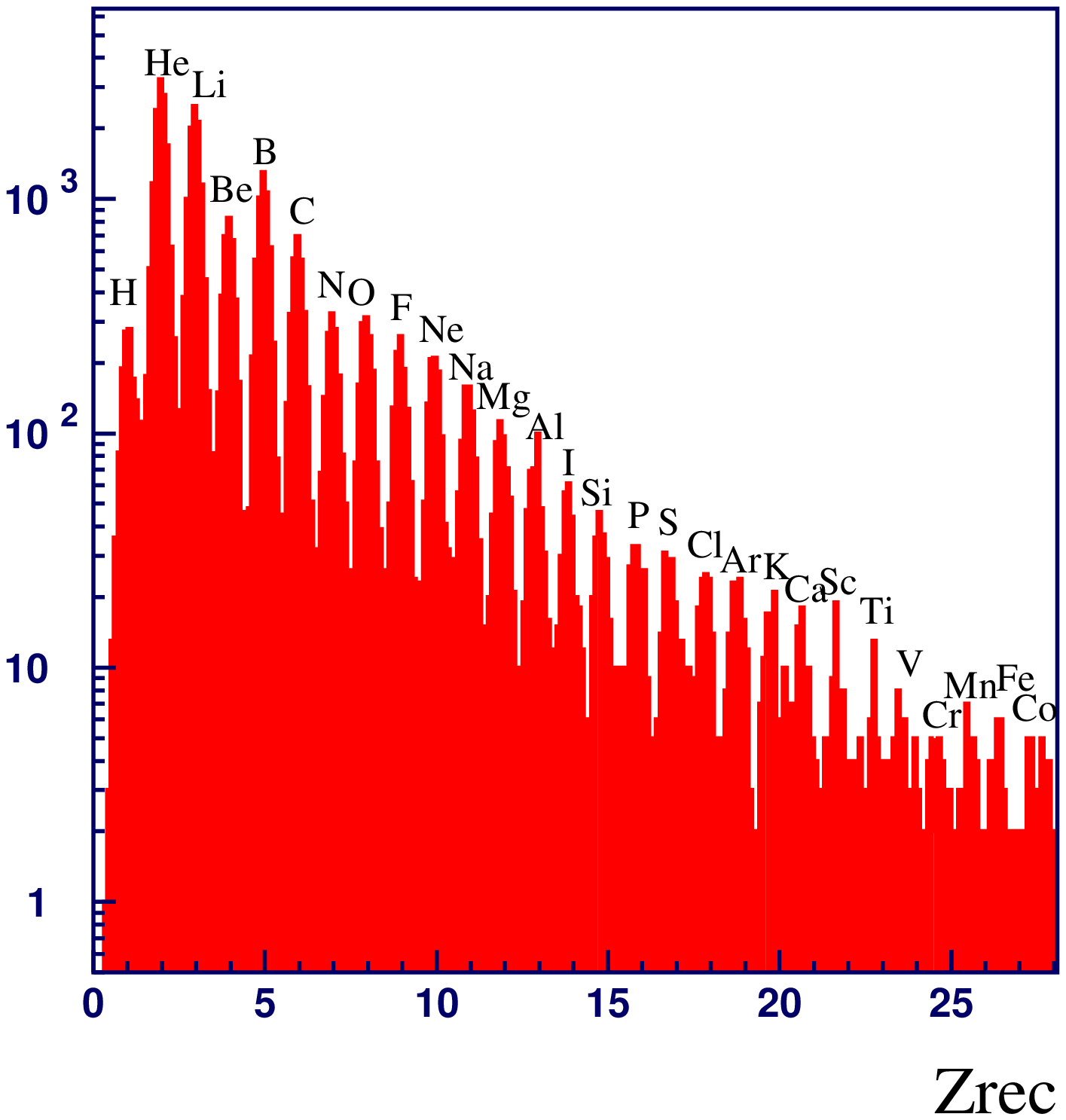,width=0.5\textwidth,clip=}}

\end{tabular}

\vspace{-0.2cm}

\caption{Results from RICH prototype tests: velocity resolution
\emph{(left)} and charge distribution \emph{(right)} for aerogel
events.\label{protores}}

\vspace{-0.2cm}

\end{figure}

\vspace{-0.3cm}

\section{RICH prototype tests}

Tests on RICH particle detection were performed using a prototype
corresponding to a fraction of the final detector (96 PMTs).
The first prototype tests were performed in mid-2002 at LPSC
in Grenoble by exposing the detector to the cosmic-ray flux at ground
level. Later, in-beam testing took place at CERN SPS in October 2002 and
October 2003 using secondary beams of nuclear fragments\cite{bib:nim2006}.
The 2002 beam was produced by bombarding a Be target with Pb ions having a
momentum of 20 GeV/c/nucleon, while the 2003 beam was obtained from the
bombardment of a Pb target with In ions having a momentum of 158 GeV/c/nucleon.
During the 2003 tests a mirror segment corresponding to a 30$^\textrm{o}$
sector (1/12 of total) was included in the experimental setup.

Fig.~\ref{protores} shows prototype results for velocity and charge
reconstruction with the aerogel radiator chosen for the final detector.
A velocity resolution $\Delta \beta / \beta = 8.7 \times 10^{-4}$ for
$Z=1$ and $\Delta \beta / \beta \simeq 0.6 \times 10^{-4}$ for $Z>20$ was
found. A good agreement between data and Monte Carlo results was observed.
Charge separation was possible up to $Z \sim 26$, with a charge resolution
$\Delta Z = 0.16$ charge units for low $Z$. Further details on velocity and
charge reconstruction can be found in Ref. \refcite{bib:icrc2007}.

\vspace{-0.3cm}

\section{System monitoring and pre-assembly testing}

A detailed monitoring of all components of the RICH detector is necessary
to ensure that systematic errors are at the level required for good velocity
and charge reconstruction. A good velocity measurement requires a knowledge
of the aerogel's refractive index with a precision of $10^{-4}$. The
quality of charge measurement constrains several variables: required
knowledge of detector properties is at the level of \mbox{0.4 mm} for
aerogel thickness, 3\% for aerogel clarity, 1\% for mirror reflectivity
and 5\% for PMT gain and unit cell efficiency.
Extensive testing has been performed on the RICH detector and its
individual components to fulfill these requirements. A detailed
mapping of each aerogel tile's thickness and refractive index was
performed at LPSC in Grenoble with participation from LIP and UNAM. Mirror
reflectivity was measured for several incidence angles. Studies for the
characterization of PMTs and unit cells took place at CIEMAT.

Additional tests focused on the response of detector components to the
conditions expected to occur during detector transportation into orbit and
operation at the ISS. One rectangular grid, corresponding to approximately
one-fifth of the detection matrix, underwent magnetic field testing at CERN
and at LCMI in Grenoble to determine if PMT response would change significantly under
a stray magnetic field of up to \mbox{300 G}. Thermal and vacuum testing, which
included individual PMTs being subject to temperatures between
\mbox{$-35 ^\textrm{o}$C} and \mbox{$55 ^\textrm{o}$C} and a rectangular
grid undergoing temperatures between \mbox{$-20 ^\textrm{o}$C} and
\mbox{$40 ^\textrm{o}$C}, took place at CIEMAT. Vibration tests of the
radiator container (including all NaF tiles and 1/4 of aerogel tiles) took
place at SERMS in Terni, while one rectangular grid underwent similar
testing at INTA in Madrid. Individual unit cells also underwent vibration
tests. Monitoring and testing will continue during the detector assembly
period.


\begin{figure}[htb]

\center

\vspace{-0.6cm}

\begin{tabular}{cc}  

\mbox{\epsfig{file=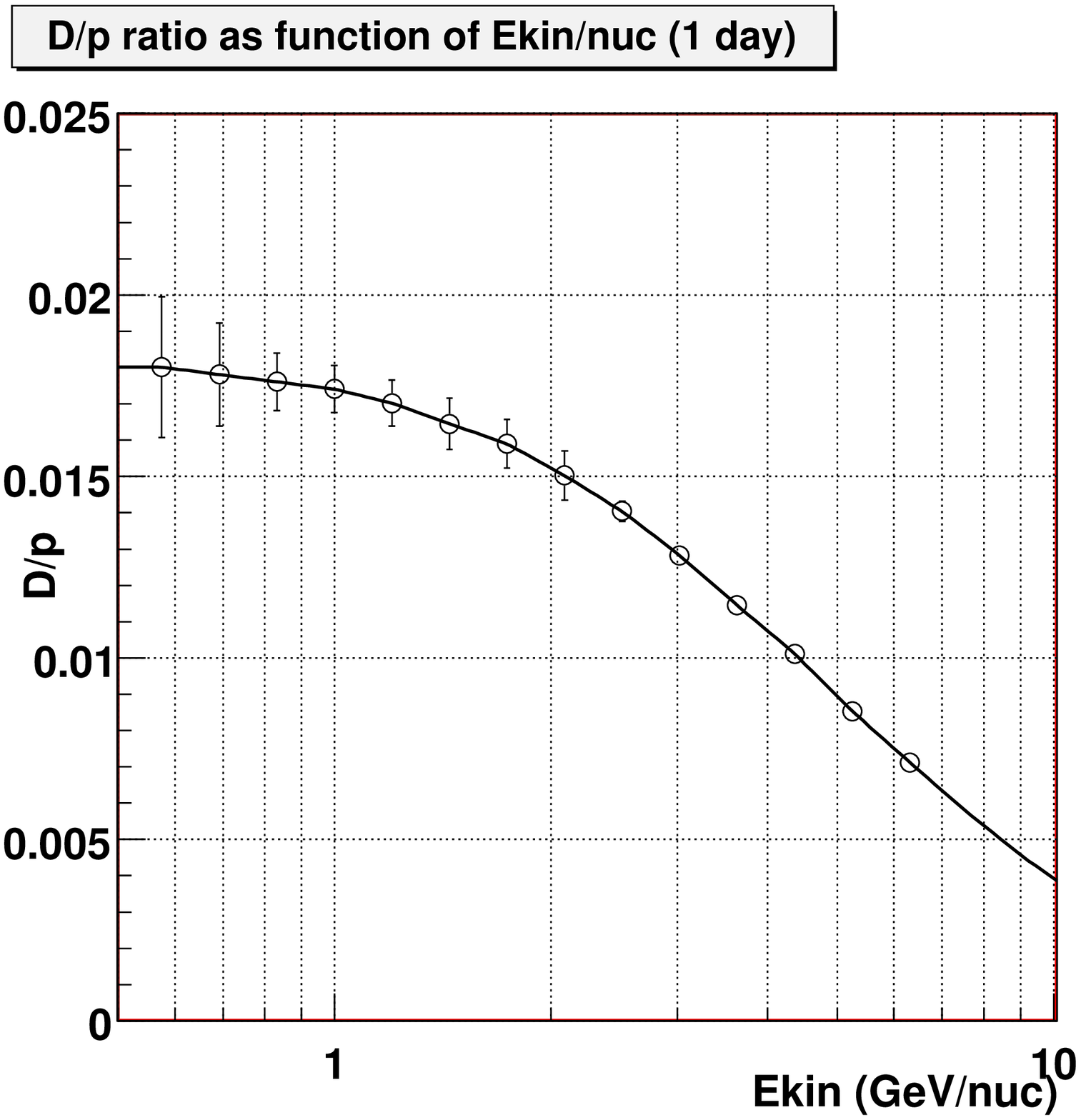,width=0.42\textwidth,clip=}}
&
\mbox{\epsfig{file=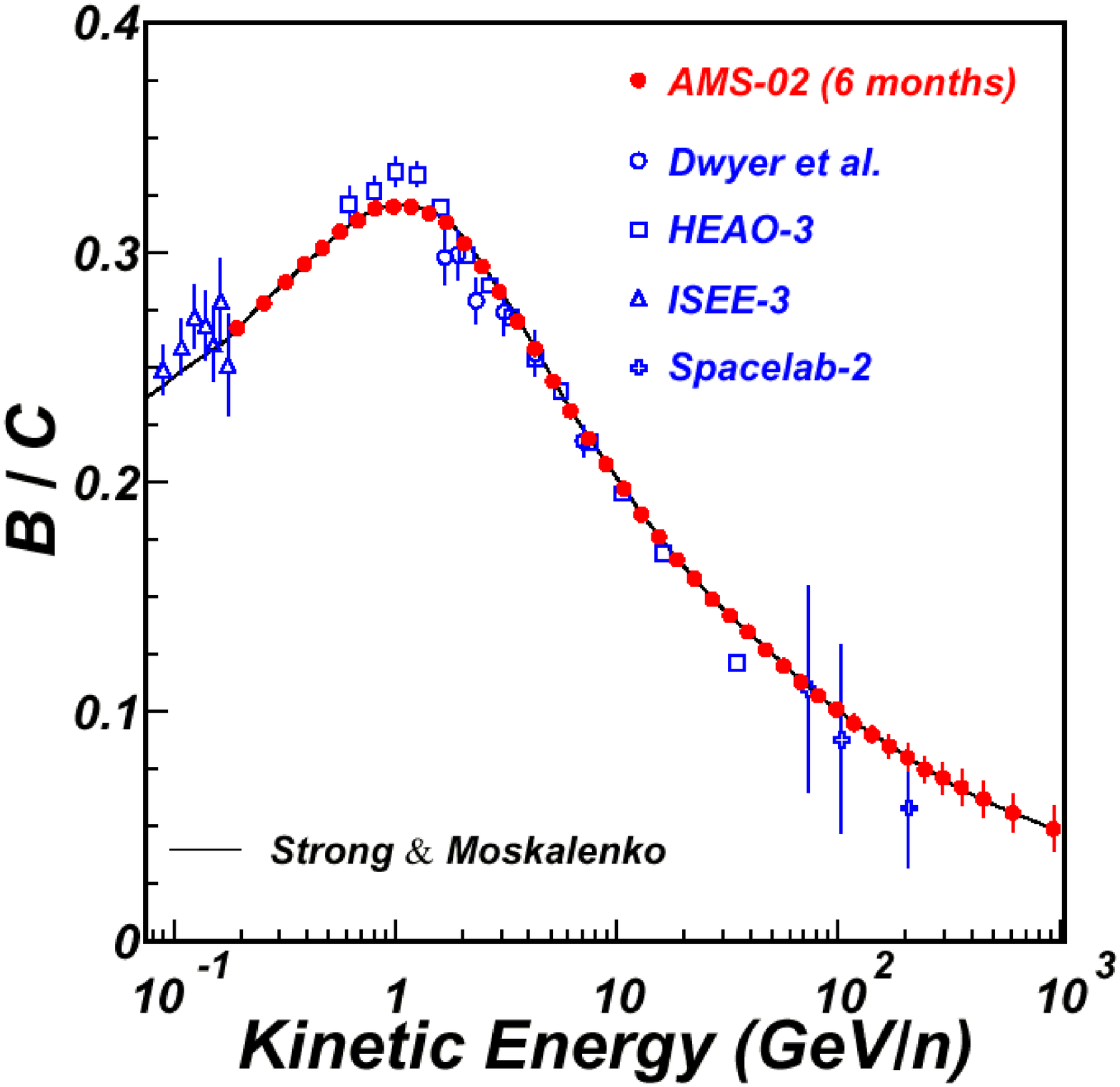,width=0.42\textwidth,clip=}}

\end{tabular}

\vspace{-0.3cm}

\caption{Expected sensitivity of AMS: D/p ratio with one day of data
\emph{(left)} and B/C ratio with six months of data
\emph{(right)}.\label{bcdpratios}}

\vspace{-0.2cm}

\end{figure}

\vspace{-0.4cm}

\section{Physics prospects}

The capacities of the RICH detector for velocity and charge reconstruction
will play a major role in particle identification with AMS-02. RICH data
will provide charge separation up to $Z \sim 26$. Mass measurements obtained
from the combination of RICH velocity with Tracker rigidity data will make
possible the isotopic separation of light elements such as H, He and Be.

AMS data will provide new insights on cosmic ray physics. Secondary to
primary ratios such as D/p, $^3$He/$^4$He and B/C will provide information
on cosmic ray propagation, while the ratio between the abundances of
radioactive isotope $^{10}$Be and stable $^9$Be will provide data on
galactic confinement times and help improve the accuracy of existing halo
models.

\vspace{-0.4cm}

\section{Conclusions}

AMS-02 will provide a new insight on the cosmic-ray spectrum by collecting
precise data for an unprecedented number of particles above the Earth's
atmosphere. The RICH detector will play a key role in the operation of AMS
due to its capabilities for velocity reconstruction, charge determination
and albedo rejection. Extensive testing has been performed on the RICH
detector and its components. The assembly of the RICH detector is currently
being finished and the full AMS detector is expected to be ready
by the end of 2008.

\vspace{-0.4cm}


\begin{thebibliography}{99}                                 

\bibitem{bib:ams}        S. P. Ahlen {\it et al.}, {\it Nucl. Instrum. Methods A} {\bf 350}, 34 (1994).\\
                         V. M. Balebanov {\it et al.}, {\em AMS proposal to DOE}, approved April 1995.

\bibitem{bib:nim2006}    P. Aguayo {\it et al.}, {\it Nucl. Instrum. Methods A} {\bf 560}, 291 (2006).

\bibitem{bib:icrc2005}    F. Barao {\it et al.}, Proceedings of 29th ICRC, Pune, India, 349-352 (2005), arXiv:astro-ph/0603852.

\bibitem{bib:icrc2007}    F. Barao {\it et al.}, Proceedings of 30th ICRC, Merida, Mexico (2007), arXiv:0709.2154 [astro-ph].








\end{thebibliography}

\end{document}